\documentclass[]{spiejour}
\usepackage{amsmath,amssymb,amsfonts,epsfig}
\usepackage[applemac]{inputenc}

\def \be{\begin{equation}}
\def \ee{\end{equation}}

\begin{document}

\title{A quantum way for metamaterials}

\author{Didier Felbacq}
\affiliation{Universit\'{e} de Montpellier II\\Laboratoire Charles Coulomb \\
Unit\'e Mixte de Recherche du Centre National de la Recherche Scientifique 5221\\
34095 Montpellier Cedex 5, France\\}
\authorinfo{Further author information: (Send correspondence to D. Felbacq)\\D. Felbacq: E-mail: Didier.Felbacq@univ-montp2.fr, Telephone: +33 (0)467 143 216}

\maketitle
\begin{abstract}
A new future for metamaterials is suggested, involving the insertion of quantum degrees of freedom, under the guise of quantum dots or cold atoms, in an photonic matrix. It is argued that new emergent, quantum, properties could be obtained.
\end{abstract}
\keywords{metamaterials, quantum mechanics, ultracold quantum gases}

The field of nanophotonics witnesses a sort of golden age, where many new fields, such as Plasmonics, Metamaterials, Transformation optics, have risen recently, leading to a wealth of new directions of research.
The story started more than 20 years ago, when new artificial structures were imagined that could exhibit a photonic band gap. These structures were called photonic crystals \cite{phc}. They have carried a lot of hope on the possibility of molding the flow of light. It was indeed soon recognised that they could go way beyond the band gap and that, in fact, they had a very rich band structure that allowed for a control over the propagation of light itself, {\it inside} the photonic crystal. This has lead to such effects as ultra-refraction, slow light, control of second harmonic emission, gradient photonic crystals, negative refraction. Somehow, this wealth of properties has slowed down researches with  the original quantum flavor of photonic crystals, at least as it appeared in the work of Sajeev John, which were seen really as devices able to control the Purcell effect. This was for instance the case for the idea of realizing a laser cavity with a vey low threshold. This aspect has known a renewed interest recently \cite{lascav}. 

If the history of nanophotonics is followed further, the concept of metamaterials is encountered \cite{meta}. It is somehow a generalization of photonic crystals, in that it is artificial periodic devices whose basic cell can be however very complicated. The point at issue is to look at these structures from the point of view of the effective properties. That is to envision these structures when they are illuminated by an incident plane wave whose wavelenght is larger than twice the period. In that situation, it can be hoped to describe the metamaterials by homogeneous constitutive relations. It is what happens in matter, at a much lower scale: for wavelengths of, say, the visible spectrum, matter, despite its discrete, granular aspect, appears as if it were homogeneous with permittivity and permeability tensors describing how it interact with light at a macroscopic level. The point at issue is of course to be able to derive these macroscopic properties from the microscopic ones. This comes under homogenization theory \cite{homog, homog2, homog3}.
Homogenization theory is quite an old subject, in the field of electromagnetics, it occupied people by the end of the $19^{th}$ century already, and during the $20^{th}$ century it was applied to mechanics (with the theory of thin plates) as it spreads towards the community of mathematicians. It is nowadays a hot topics, precisely due to the interest in the effective properties of metamaterials. Here a parenthesis should be opened on the lack of communication between physicists and mathematicians. If one browses the now vast literature of physics on the effective properties of metamaterials, one notes that there is almost no citations of any of the numerous mathematicians involved in the rigorous treatment of the theory (I have in mind the french school with Lyons, Bensoussan, Tartar, Murat, Allaire, Bouchitt\'e, Villani, the russian school: Jikov, Oleinik, Kozlov, the italian school: De Giorgi, Butazzo, the american school: Papanicolaou, Vogelius and so on, many names being omitted). Despite the fact that many excellent books were written, describing very clearly the theory \cite{kozlov}, it seems that very few physicists are aware of the very existence of this theory. Rather, the physics literature seems to be a vast collection of recipes, full of acronyms, with no results duly proven or any clear theorem stated. It is not that this computational approach is not with any interest, but it lacks the conceptual clarity of stating something like: when this small parameter tends to 0, the field tends (in some precise meaning, e.g. in $L^2$) towards the solution of such and such PDE\footnote{It should be added for fairness that mathematicians should sometimes try to put their results in a form that is usable to physicists.}.

Anyway, the point at issue here is that metamaterials are for the moment purely classical structure: they are made of wires and metallic loops, dielectric or metallic spheres or nanorods. They do not belong for the moment to the quantum world, although it should be noted that the properties of photonic structures in which quantum systems are embedded have been already explored \cite{quantmat}. 
The point at issue in what could be called "quantum metamaterials" would be to use quantum-in-essence systems (e.g. molecules, quantum dots, cold atoms, Josephson junctions) embedded in a photonic structure in order to tailor the effective properties. Here "effective properties" could mean not only electromagnetic constitutive relations, but, more generally, emergent collective behaviors\footnote{in that respect superconductors are somehow metamaterials!}. For instance it was already suggested in \cite{zagoskin} (using the term "quantum metamaterials") to use arrays of Josephson junctions to exploit quantum coherence in metamaterials. More generally, it could be possible to reproduce the full range of atomic spectroscopy spectra.
\begin{figure}
\begin{center}
\includegraphics[width=8cm]{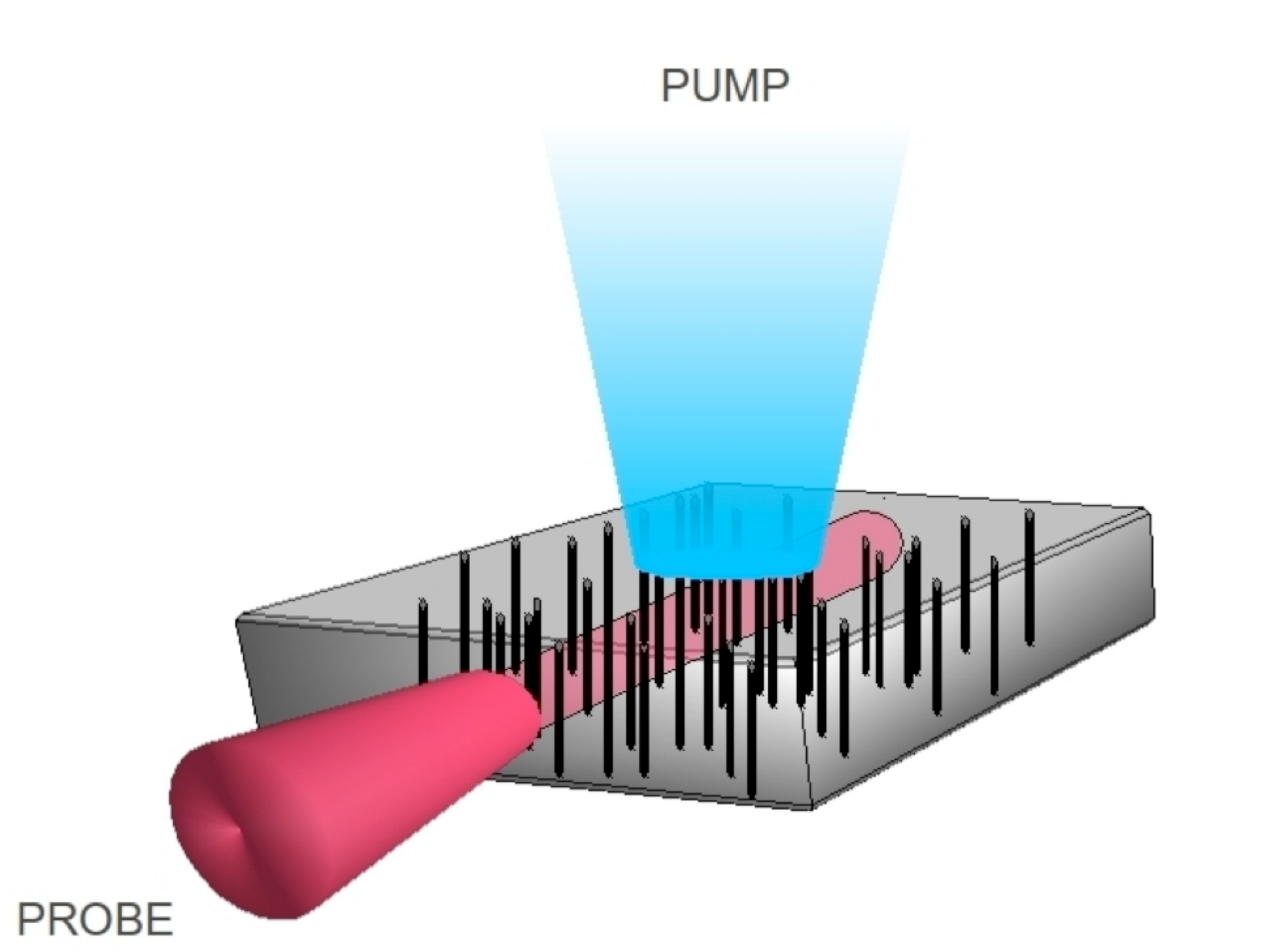}
\caption{Sketch of a quantum metamaterials based on nanowires with quantum dots inserted. The propagation of the probe beam can be tuned by the pumped beam.}
\end{center}
\end{figure}
 As compared to what happens in the basic elements of metamaterials (i.e. open cavity resonances), what would be most interesting would be to be able to address the quantum dynamics that allows for transitions between levels (inversion of population), the entanglement between states (quantum information applications), the tunability the effective properties. Let us give an example as a first step: consider an array of nanowires such as that depicted in fig.1. They can be put into periodic positions with some lattice symmetry, so as to produce a photonic band gap. It is possible to grow inside the wire a quantum dot with an electric dipole perpendicular to the axis of the wire. By pumping this medium it is possible to make a transition from a conducting state to a band gap \cite{moicas}.
 
Finally, I would like to sketch an interesting connection between cold atoms and metamaterials. The limitations of the model of an artificial atom for the description of quantum dots have been analyzed \cite{cass} and the existence of phonons affects the emission properties. I would like to stress that periodic arrays of cold atoms could be an ideal model of a solid-state quantum metamaterials, i.e. quantum dots embedded in a photonic crystal (or a classical metamaterial). Indeed, quantum dots are never alike and the size dispersion can make collective effects disappear but atoms are all identical. Somehow, ultracold gases are already considered as quantum metamaterials: they allow to reproduce the many-body physics of solid-state physics \cite{dalibard}.   
\acknowledgments{I thank G. Cassabois and M. Antezza for enlightning discussions. The financial support of the Institut Universitaire de France is gratefully acknowledged.}

\end{document}